\documentclass[aps,pre,reprint,amsmath,amsfonts,superscriptaddress]{revtex4-1}
\usepackage[english]{babel}
\usepackage{times}

\usepackage{tikz}
\definecolor{gray}{cmyk}{0,0,0,0.4}
\definecolor{lightgray}{cmyk}{0,0,0,0.08}
\tikzstyle{trans}=[->,line width=1.5pt,shorten <=0.5em,shorten >=0.5em]
\tikzstyle{aplayer}=[circle,draw=black,thick,inner sep=0pt, minimum
size=1.3em,fill=lightgray]
\tikzstyle{bplayer}=[circle,draw=black,thick,inner sep=0pt, minimum
size=1.3em,fill=gray]
\tikzstyle{every child}=[node distance=10em,level distance=3em, sibling
distance=3em]
\tikzstyle{edge from parent}=[-,thick,draw]
\tikzstyle{every label}=[inner sep=0pt]

\begin{document}

\title{Largenet2: an object-oriented programming library for simulating large
adaptive networks}
\author{Gerd Zschaler}
\email{gerd@biond.org}
\affiliation{Max-Planck-Institut f\"ur Physik komplexer Systeme, N\"othnitzer
Str. 38, 01187 Dresden, Germany}
\author{Thilo Gross}
\affiliation{Department of Engineering Mathematics, Merchant Venturers Building,
University of Bristol, Woodland Road, Clifton, Bristol BS81UB, UK}

\begin{abstract}
The largenet2 C++ library provides an infrastructure for the simulation of large
dynamic and adaptive networks with discrete node and link states.
The library is released as free software. It is available at
http://rincedd.github.com/largenet2.
Largenet2 is licensed under the Creative Commons Attribution-NonCommercial 3.0
Unported License.
\end{abstract}

\maketitle

The investigation of dynamical processes on networks has become a highly active
research field, which addresses questions from a wide range of disciplines
\citep{Barrat2008, Newman2010}. One of the fundamental tools of
``network science'' \citep{Boerner2007} is computer simulation. Prominent
examples include the study of the
propagation of communicable diseases in networks of social contacts
\citep[e.g.][]{Kuperman2001}, the emergence of consensus in networks of
interacting agents \citep{Castellano2005, Sood2005}, or the evolution of
cooperation among selfish individuals \citep{Santos2005, Nowak2006}.

Over the past decade in particular, \emph{adaptive} networks have received a lot
of attention. In this class of network models the network structure itself
changes dynamically in response to the dynamics of its constituents
\citep{Gross2008, Gross2009}. This creates a feedback loop between the dynamics
on the network and the dynamics of the network itself, leading to emergent
complex behaviour. For instance, adaptive-network models have been
studied for social networks \citep{Skyrms2000}, opinion
formation \citep{Vazquez2008, Nardini2008, Durrett2012}, epidemic
spreading \citep{Gross2006, Shaw2008}, and collective motion \citep{Huepe2011,
Couzin2011}.

\begin{figure}\centering
 (a)\hspace{0.5em}\begin{tikzpicture}[baseline=(one.base)]
   \path (0, 0) node[aplayer] (one) {S};
   \path (1, 0) node[bplayer] (two) {I};
   \draw[-,thick] (one) -- (two);
   \path (2.5, 0) node[aplayer] (one1) {I};
   \path (3.5, 0) node[aplayer] (two1) {I};
   \draw[-,thick] (one1) -- (two1);
   \draw[trans] (two) -- (one1) node[midway,yshift=1.5ex] {$p$};
\end{tikzpicture}\hspace{4em}
 (b)\hspace{0.5em}\begin{tikzpicture}[baseline=(one.base)]
   \path (0, 0) node [bplayer] (one) {I};
   \path (1.5, 0) node [aplayer] (one1) {S};
   \draw[trans] (one) -- (one1) node[midway,yshift=1.5ex] {$r$};
\end{tikzpicture}\bigskip

 (c)\hspace{0.5em}\begin{tikzpicture}[baseline=(one.base)]
\tikzstyle{edge from parent}=[]
\node[aplayer] (one) {S} [grow'=right]
	child {node[bplayer] (two) {I}}
	child {node[aplayer,dashed] (three) {S}};
\draw[-,thick] (one) -- (two);
\path (two.south) -- (three.north)  node[midway,aplayer,fill=none,draw=none] (m1) {};

\node[aplayer,xshift=2.5em] (one1) [right of=m1] {S} [grow'=right]
	child {node[bplayer] (two1) {I}}
	child {node[aplayer,dashed] (three1) {S}};
\draw[-,thick] (one1) -- (three1);
\draw[trans] (m1) -- (one1)
	node[midway,yshift=1.5ex] {$w$};
\end{tikzpicture}
 \caption{\label{fig:rules}Diagrammatic representation of the transformation
rules for an epidemiological model  \citep{Gross2006}. (a) a susceptible node
(S) is infected through its link to an infectious neighbour (I) with probability
$p$ per S-I-link and unit time; (b) an infectious node (I) recovers with
probability $r$; (c) a susceptible node (S) breaks its connection to an
infectious neighbour (I) and rewires to another randomly selected susceptible
node (dashed) with probability $w$. This last rule makes the network adaptive,
because it changes the topology depending on the node states.}
\end{figure}
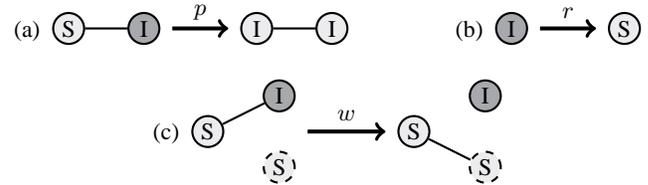

Dynamical processes in adaptive networks are typically specified in
terms of a set of rules that locally transform a part of the network, e.g.,
update a node's state according to its neighbourhood or modify the local
connectivity of a node \citep{Zschaler2012thesis, Gorochowski2012}.
An example of such rules for
an epidemiological model is shown in Figure~\ref{fig:rules}. The transformation
rules can be directly implemented in computer simulations. For stochastic
models, they are typically applied asynchronously using Monte Carlo techniques
such as Gillespie's algorithm \citep{Gillespie1976}.

In order to apply the transformation rules efficiently in simulations, the
network subgraphs involved in a specific rule must be accessible at random,
i.e., they must be located directly without resorting to an extensive search in
the network. For instance, for the infection rule in Fig.~\ref{fig:rules}~(top),
efficient access to the links connecting S- and I-nodes in the network must be
provided. Thus appropriate data structures representing the network are required
which provide random access to the network nodes, links, and similar subgraphs,
store properties such as node and link states, and allow for fast changes of the
network topology.

Standard data structures used to represent networks (or graphs) in computer
science are tailored towards the efficient implementation of certain algorithms,
as for instance graph traversal, search, or finding shortest paths
\citep{Sedgewick2002, Siek2002}. In most cases, these algorithms work on static
networks with a fixed topology, and efficient access to node and link
states is usually not of major concern. Such data structures are
therefore not suitable for the the simulation of large adaptive networks, whose
structure changes dynamically and depending on the node and link states.

The \textit{largenet2} library has been developed specifically for the
efficient simulation of dynamic and adaptive networks. It provides data
structures for networks
with discrete node and link states (represented as integer numbers), allowing
for fast random access to nodes and links in any given state, and efficient
manipulation of these states and the network topology. Nodes, links, and, if
required, larger subgraphs are stored in a custom-made, index-based container
which can hold items in different discrete
categories (states). It ensures that items in the same category are stored in
contingent memory and provides both index-based and category-based access, so
that selecting a random item in a given category can be achieved in constant
time.

The network structure is modelled directly in memory using
nodes and links as the basic entities in a double adjacency set representation,
in which each node keeps a set of pointers to its incoming and
outgoing links. At the cost of some memory overhead, addition and removal of
links is thus achieved in logarithmic time.
In effect, simulating large adaptive networks with \textit{largenet2} is
typically of linear complexity, i.e., the required simulation time scales
linearly with the number of nodes in the network.

Additionally, the \textit{largenet2} library
provides a basic stochastic simulation framework implementing the original
direct method of Gillespie's algorithm \citep{Gillespie1976} and one of its
variants \citep{Allen2009}.
The library consists of the following main packages, organized in different
namespaces:
\begin{itemize}
 \item network data structures for directed or undirected networks with or
without parallel links (\texttt{largenet})
 \item generation of random networks (\texttt{largenet::generators})
 \item basic network measures, such as degree distributions and correlations
(\texttt{largenet::measures})
 \item network file input/output of edge list files and other file formats
(\texttt{largenet::io})
 \item stochastic simulation (\texttt{sim::gillespie})
\end{itemize}

For
implementation details, examples, and source code documentation, please refer to
the website.

The \textit{largenet2} library and its predecessor \textit{largenet} have been
used for the simulations of large adaptive networks in \citep{Zschaler2010,
Boehme2011, Demirel2011, Zschaler2012}. To implement more complex transformation
rules than depicted in Fig.~\ref{fig:rules}, the library can be and has
been extended to also track larger network subgraphs, such as node triplets,
involved in such rules \citep[e.g.][]{Couzin2011, Huepe2011}.

The open source library \textit{largenet2} is under ongoing development. It is
set up as a community effort and contributions are welcome at
http://github.com/rincedd/largenet2.

\bibliographystyle{natbib}

\end{document}